\documentstyle[prd,aps,epsfig]{revtex}
\begin{document}
\draft
\title{Maximally localized states and causality in non commutative quantum theories}
\author{Musongela Lubo\footnote{email muso@umhsp02. umh. ac.
be}$\mbox{}^{\mbox{\footnotesize }}$,\\
         \vskip0.25cm
          M\'ecanique et Gravitation, Universit\'e de Mons--Hainault,\\
             6 avenue du Champ de Mars, B--7000 Mons (Belgium)}
\maketitle
\begin{abstract}
We give simple representations for quantum theories in which the
position commutators are non vanishing constants. A particular
representation reproduces results found using the Moyal star
product. The notion of exact localization being meaningless in
these theories, we adapt the notion of ``maximally localized
states'' developed in another context . We find that gaussian
functions play this role in a $2+1$ dimensional model in which the
non commutation relations concern positions only. An
interpretation of the wave function in this non commutative
geometry is suggested. We also analyze  higher dimensional cases.
A possible incidence on the causality issue for a Q.F.T with a non
commuting time is sketched.
\end{abstract}
\pacs{{\bf PACS numbers:03.65-w}} \vskip2pc
%%%%%%%%%%%%%%%%%%%%%%%%%%%%%%%%%%%%%%%%%%%%%%%%%%%%%%%%%%%%%%%%%%%
\section{Introduction}\label{sec_intro}
 Non commutative quantum mechanics have received a wide
attention once it was realized that they could be obtained as low
energy limits of string theory in the presence of a B field
\cite{LABEL1,LABEL2}. However, the status of these theories is
still plagued by conceptual challenges. In most of them, the
Lorentz invariance is explicitly broken. Actually, non commutative
quantum mechanics are not the only arena in which the Lorentz
symmetry is only approximate. For example, it has been suggested
recently that the standard model itself may fit into this category
\cite{LABEL3,LABEL4}. The preferred frame was postulated to be the
rest frame of the cosmic microwave background radiation. There is
one major problem quantum non commutative theories  are believed
to  face when time and position do not commute: the lack of
causality and unitarity \cite{LABEL6}. The analysis which led to
this result relies on the Weyl-Moyal correspondence which tells us
how to handle theories with non commuting positions. One simply
works with functions of commuting variables but replaces any
pointwise product by the Moyal star product which is non local.

\par One of the main problems faced by this approach is that  the meaning of the wave function
is not clarified yet \cite{LABEL7,LABEL8}. When two positions  do
not commute, they can not be diagonalized simultaneously. The
uncertainty relation prevents the wave function
  $\phi(\vec x)$ appearing in non commutative theories from being the probability for
  a particle to be localized at $\vec x$.

\par There is a model in which exact localization is also forbidden:
the K.M.M(Kempf-Mangano -Mann)
 theory \cite{LABEL9}. Inspired by what was done in this case, and in \cite{LABELD}we will
 adapt the notion
 of ``maximally localized state'' to non commutative quantum
 mechanics.This notion will be useful in the discussion of the
 causality issue.
\par The plan of the paper is as follows. In the second section we
will rapidly point out some characteristics of non commutative
quantum mechanics which have been obtained in specific cases using
the Weyl-Moyal correspondence \cite{LABEL10,LABEL11}. In the third
section we will exhibit a representation of the positions and the
momenta by operators acting on a usual space of functions. We will
show that the results summarized in section two are recovered. In
the fourth section, we first work in a 2+1 dimensional model where
the spatial coordinates do not commute. We give all the
 details leading to gaussian functions as being  maximally localized states. Considering
  a 3+1 dimensional theory, we use the preceding construction to
  construct the appropriate states in the new context.
\par Rather than projecting on localized states, we now
   have to project on maximally localized states to gain viable information on positions.
   The last section is devoted to a brief reminder of the causality issue of a Q.F.T possessing
a non commuting time. The way maximally localized states may alter
the analysis is sketched.

\section{Quantum mechanics using the Moyal product.}
\par The non commutative quantum theories we are interested in obey the following relations
\begin{equation}
\label{1}
 [\hat x_{k}, \hat x_{l}] = i\theta_{kl}\quad ,
\quad [\hat p_{k}, \hat p_{l}] = 0\quad , \quad [\hat x_{k}, \hat
p_{l}] = i\hbar \, \delta_{kl}.
\end{equation}
The constant matrix $\theta$ has dimension $L^2$ and breaks
explicitly the Lorentz invariance. The Weyl-Moyal correspondence
is a map between the functions $\Phi$ of the operators $\hat x_k$
and the functions $\phi$ of the commuting variables $x_i$ :
\begin{equation}
\label{2}
 \Phi(\hat x) = \int e^{i\alpha.\hat x} \tilde
\phi(\alpha) d\alpha \quad , \quad  \phi(x) = \int e^{- ix\beta}
\tilde \phi (\beta) d\beta \quad .
\end{equation}
The usual product of two $\hat x$ valued functions is sent to the
star  product of the associated functions defined on commuting
variables:
\begin{equation}
\label{3} \Phi (\hat x) \Psi (\hat x) \longrightarrow (\phi \ast
\psi) (x) \quad ,
\end{equation}
with
\begin{equation}
\label{4}
 (\phi \ast \psi) (x) = \left[ e^{{i\over
2}\theta^{\mu\nu}\partial_{\xi^{\mu}}
\partial_{\eta^{\nu}}} \phi(x+\xi) \psi(x+\eta) \right]_{\xi=\eta=0}  \quad .
\end{equation}

The theories with non commuting positions are obtained from the
usual actions in which all products become star products. For
example, the action of a self interacting scalar field reads
\begin{equation}
\label{5}
 S = \int d^4x \left( \frac{1}{2}g^{\mu\nu} \partial_{\mu}\phi \ast
\partial_{\nu}\phi - m^2\phi\ast \phi - {\lambda\over {4!}} \phi \ast
\phi \ast \phi \ast \phi \right).
\end{equation}
It has been suggested that quantum mechanics could be derived in
this context by a similar replacement \cite{LABEL10,LABEL11}. In
the usual situation, the Schr\"odinger equation can be inferred
from the action
\begin{equation}
\label{6}
 S = \int dt \, d^2x \bar \psi \left[i \partial_t - {\vec
p^2\over {2m}} - V(x) \right]\psi
\end{equation}
in a 2+1 dimensional system. Introducing star products and using
the relation
\begin{equation}
\label{7}
 \label{124} V(x) \ast \psi (x) = V \left( x-{\tilde
p\over 2 }  \right) \psi(x)
\end{equation}
(where $\tilde p^i = \theta^{ij} p_j$) which is obtained  via a
Fourier transform \cite{LABEL10,LABEL11}, one finds that the
Weyl-Moyal correspondence  reduces to the replacement
\begin{equation}
\label{125} x\rightarrow x - {1\over 2} \tilde p
\end{equation}
 in the potential.
 Considering a central potential, the substitution
\begin{equation}
\label{126} V( x^2+y^2) \rightarrow V \left( {\theta^2\over 4}
p^2_x + {\theta^2\over 4} p^2_y + x^2+y^2-\theta L_z \right)
\end{equation}
shows that the theory ``looks like''  one describing a particle of
changed mass, with a non trivial coupling to the angular momentum
\cite{LABEL10}. In the case of an asymmetric oscillator $V(x,y) =
{1\over 2} ky^2$, the theory contains a term which looks like an
interaction between a particle of charge $q$  with a constant
magnetic field $B$ such that $q B = {2\over {\theta}}$
\cite{LABEL11}.
  At this point it is crucial to realize that although the Moyal
  product is written in terms of commuting variables $x$, these
  variables are simply a notation. There is no evidence that they represent physical
  coordinates,except in the undeformed case $ \theta = 0$. No argument has been presented
  which shows that the wave function $ \phi $
  gives a probability \cite{LABEL7}. Moreover, the probability for
  a particle to be localized at a given position $ (x_1,x_2) $ is
  not a safe concept since these  coordinates do not commute.

\section{A representation of the commutation relations.}\label{sec_repr}

The only modification to the usual theory introduced by
Eq.(\ref{1}) concerns  the positions. It is therefore quite
reasonable to look for a realization in which the momenta remain
unchanged :
\begin{equation}
\label{127} \hat p_i = -i\hbar \partial_{\xi_i} \quad .
\end{equation}
The introduction of the non commutativity scale leads to the
possible ens\"atze
\begin{equation}
\label{128} \hat x_{i} = \xi_{i} + \theta^{1/2} G_i( \theta^{1/2}
\partial_{\xi^{k}})  \quad .
\end{equation}
The functions $ G_i $ are taken  analytic. Such an ensätz clearly
fulfills the $[\hat x,\hat p]$ commutation relations. The
$[\hat{x}, \hat{x}]$ commutators can then be used to constrain the
coefficients of the Taylor expansions of the functions $G_k$.
\par As an illustration, let us consider a 2+1 dimensional system, with spatial non
commutativity :
\begin{equation}
\label{129} \quad [\hat x_1, \hat x_2] = i\theta \quad ,
\end{equation}
with $ \theta $ positive. It is straightforward that one can take
$G$ linear and write simply
\begin{eqnarray}
\label{130}
\hat x_1 &=& \xi_1 + i\theta  \, (a \, \partial_{\xi_1} + (1+c) \, \partial_{\xi_2}) \quad , \nonumber\\
\hat x_2 &=& \xi_2 + i \, \theta (c \,
\partial_{\xi_1}+ d \, \partial_{\xi_2}) \quad .
\end{eqnarray}
At this stage, the constants $ a,c$ and $d $ are arbitrary. The
momenta $\hat p_i$ and the positions $\hat x_k$ act as operators
on the space of functions of the variables $(\xi_1,\xi_2)$. If we
take the scalar product to be given by the usual formula
\begin{equation}
\label{131} \langle\phi|\psi\rangle = \int d\xi_1  \,\, d\xi_2
\,\, \phi^{\ast}(\xi_1, \xi_2) \,\, \psi(\xi_1, \xi_2) \,\,\,  ,
\end{equation}
the $\hat x_i$ operators are symmetric provided that $a,c$ and $d$
are real.
\par Let us consider a harmonic oscillator in this theory :
\begin{equation}
\label{132} \hat H = {1\over {2m}} (\hat p^2_x + \hat p^2_y) +
{1\over 2}k(\hat x^2 + \hat y^2) \quad .
\end{equation}
Using the representation given above, one obtains
\begin{eqnarray}
\label{133} \hat H = &-& \left({\hbar^2\over {2m}} + {1\over 2} k
\theta^2(a^2+c^2)\right)\partial^2_{\xi_1} -
\left({\hbar^2\over{2m}} +
{1\over 2} k\theta^2((1+c)^2+d^2)\right) \partial^2_{\xi_2}\nonumber\\
&-& k  \theta^2  \left( a(1+c)+c d \right
)\partial_{\xi_1}\partial_{\xi_2} \nonumber\\
&+& {1\over 2} k i \theta  ( 2a \xi_1\partial_{\xi_1} + 2d
\xi_2\partial_{\xi_2}
 + 2(1+c) \xi_1\partial_{\xi_2} + 2c \xi_2
\partial_{\xi_1}+a+d ) \nonumber\\
&+& \frac{1}{2} k ( \xi_1^2 + \xi_2^2 ) \quad .
\end{eqnarray}
Let us forget for a moment the origin of this operator and treat
it like in the usual , commutative theory. Can we reproduce the
features shown in the last section, which come from an analysis
based on the Moyal product? The answer is positive. The appearance
of the `` angular momentum operator " of the usual theory $i(\xi_1
\partial_{\xi_2}-\xi_2
\partial_{\xi_1})$, is guaranteed by the choice $ c = -1/2 $ . The
crossed derivative $ \partial_{\xi_1} \partial_{\xi_2} $ vanishes
if the relation $ a = d$ holds and the terms
 $ \xi_1 \partial_{\xi_1}$ and $ \xi_2 \partial_{\xi_2}$
 disappear
 if we also impose $ a = 0 $.
The final hamiltonian reads
\begin{equation}
\label{134} \hat H = - \left( {\hbar^2\over{2m}} + \frac{1}{8} k
\theta^2 \right)
\partial^2_{\xi_1} - \left( {\hbar^2\over{2m}} + \frac{1}{8} k
\theta^2 \right)
\partial^2_{\xi_2} + {1\over 2} k \theta L + {1\over 2}k
(\xi^2_1+\xi^2_2).
\end{equation}
    This has been  obtained in $ \cite{LABEL10} $ and summarized in
    the previous section.

\par So, for a particular choice  of the free parameters appearing in the
realization we choosed for the non commutative quantum theory, we
can  reproduce exactly some results derived using the Weyl-Moyal
correspondence. The simplicity of the algebra will prove useful
when tackling the interpretation of the wave function in this
framework. In fact, even if the hamiltonian written in
Eq.(\ref{134}) looks quite ordinary, one should keep in mind that
once the wave equation is solved, the position operator along the
first spatial coordinate is not simply the product by $\xi_1$. The
energy eigenvalues have the same meaning than in the ordinary
theory but the analysis concerning localization is much more
involved. A similar situation occurs in another theory and has
been exploited to handle the transplanckian problem of the black
hole physics \cite{LABEL12,LABEL13}.
\par From the formula given in Eq.(\ref{1}), one infers the uncertainty relation
\begin{equation}
\label{135} \Delta x_1 \Delta x_2 \geq {\theta\over 2} \qquad .
\end{equation}
\par This means that any state which is localized without any uncertainty in any of the two
 directions is unphysical.

\section{Maximally localized states.}

\subsection{The derivation}
\par  The uncertainty relation given in Eq.(\ref{135}) puts a lower bound on
localization. We shall look for states which saturate this bound.
We will restrict ourselves to those displaying equal values of the
uncertainties in the two directions:
\begin{equation}
\label{8}
 \Delta x_1 = \Delta x_2 = \sqrt{\theta\over 2} \quad .
\end{equation}
This is motivated by the opinion that a state displaying a very
small uncertainty in one direction and a large one in the
remaining direction is undesirable; we adopt here a  democratic
treatment of the two coordinates.
\newline {\em We will say that a state
is maximally localized at $(\lambda_1, \lambda_2)$ if it satisfies
the equality of Eq.(\ref{8}) and if $\langle x_i \rangle =
\lambda_i$ \quad .}

\par These states are quite close to the coherent states in the usual
  quantum mechanics which verify $ \Delta x \Delta p = \hbar/2 $.
As it stands, Eq.(\ref{8}) is hardly tractable. The procedure we
shall use is directly inspired by \cite{LABEL9} and replaces these
integral equations by a differential one. The uncertainty relation
of Eq.(\ref{135}) is obtained as a consequence of the inequality
\begin{equation}
\label{139} \left| \left|  \left(\hat x_1-\langle x_1 \rangle + {\langle[\hat
x_1, \hat x_2]\rangle\over{2(\Delta x_2)^2}} (\hat x_2-\langle x_2
\rangle)\right) |\phi\rangle \right| \right|  \geq 0  \quad .
\end{equation}
The vector whose norm is considered in the preceding formula vanishes for the states $ \vert \phi \rangle $
which minimize the product of uncertainties.
 Using our expressions of the position operators, this is
converted into a partial
 differential equation for the maximally localized states. We introduce the complex coordinates
  $(u_1,u_2)$ by
\begin{eqnarray}
\label{140}
u_1 &=& (\alpha_1 + i \beta_1) \xi_1 + (\gamma_1+i \delta_1) \xi_2  \quad , \nonumber\\
u_2 &=& (\alpha_2 + i \beta_2) \xi_1 + (\gamma_2 + i \delta_2)
\xi_2 \quad ,
\end{eqnarray}
where the following constants have been introduced to simplify
future formula
\begin{eqnarray}
\label{141}
\alpha_1 &=&  {c\over D} \quad , \quad \beta_1 =  {a\over D} \quad , \quad
 \gamma_1 = {d\over D} \quad , \quad \delta_1 = {1+c\over D} \quad , \nonumber\\
\alpha_2 &=& -{d\over D} \quad , \quad  \beta_2 = {1+c\over D} \quad , \quad
 \gamma_2 = {c\over D} \quad , \quad \delta_2 = - {a\over D} \quad , \nonumber\\
D &=& -1-a^2-2c-2c^2-d^2 \quad .
\end{eqnarray}
One finds that the general solution to the partial differential
equation is
\begin{eqnarray}
\label{142}   & & \psi^{ml}_{\lambda_1, \lambda_2}(\xi_1,\xi_2)   =
f(u_2)  \nonumber\\
 & & \exp \left\lbrace {1\over {2\theta}} (2c+1-ia+id)u^2_1 -
    {1\over {\theta}} (d+a+i) u_2u_1 + {1\over {\theta}}
(\lambda_1+i \lambda_2) u_1\right\rbrace
\end{eqnarray}
with $f$ an arbitrary function. A this level an important
constraint comes from the fact that as $\theta\rightarrow 0$, we
should reobtain usual quantum mechanics. The maximally localized
states must in this limit co\"\i ncide with position eigenstates
which are delta functions. In formula, one should have
\begin{equation}
\label{143} \psi^{ml}_{\lambda_1,\lambda_2} (\xi_1, \xi_2)
\rightarrow \,\, \delta(\xi_1- \lambda_1) \,\,
\delta(\xi_2-\lambda_2)
\end{equation}
when $\theta \rightarrow 0 $.
 In distribution theory one knows that the
 Dirac delta can be expressed as the limit of some appropriate
 functions, for example
\begin{equation}
\label{144} \frac{1}{ \sqrt{2 \pi \theta}  } \exp{ \left( -
\frac{x^2}{ \theta} \right)} \,\,\, , \frac{\sqrt{\theta}}{\pi}
\frac{ \sin^2{( x/\sqrt{\theta}})}{x^2} \,\,\, ,  \frac{1}{\pi}
\frac{\sqrt{\theta}}{x^2 + \theta} \quad .
\end{equation}
  Any combination of these functions with appropriate coefficients
  tends to the delta distribution. It can be conjectured that a
  maximally localized state may just be such a combination. Our expression for
   $\psi^{ml}_{\lambda_1, \lambda_2}(\xi_1,\xi_2)$ given in
Eq.(\ref{142})  involves exponentials; this makes more reasonable
to focus on
 the  first element of the
  previous list .
Our aim is  to see if the function $f(u_2)$ can be chosen so that
the maximally localized state   is proportional to
\begin{equation}
\label{145} \exp \left\lbrace - {(\xi_1-\lambda_1)^2\over
{\sigma^2_1\theta}} - {(\xi_2-\lambda_2)^2\over{\sigma^2_2\theta}}
\right\rbrace  \quad .
\end{equation}

\par The answer is that this can be done only when the constants
appearing in the Eq.(\ref{130}) satisfy the relations $a=d=0$. To
obtain an expression which looks like Eq.(\ref{145}), one needs
the function $f(u_2)$ to be quadratic. As the expression of
$\psi^{ml}_{\lambda_1,\lambda_2}(\xi_1,\xi_2)$ given in
Eq.(\ref{142}) involves complex quantities, we choose
\begin{equation}
\label{146} f(u_2) = N \exp \left\lbrace {1\over {\theta}}
(A_1+iA_2)u^2_2 + {1\over{\sqrt{\theta}}} (B_1+iB_2)u_2\right
\rbrace \quad ,
\end{equation}
the $A_i , B_i$ being dimensionless real constants. We  separate
the real and the imaginary parts in the expression of the
maximally localized state:
\begin{eqnarray}
\label{147} \psi^{ml}_{\lambda_1,\lambda_2} (\xi_1, \xi_2) &=& N
\exp \left\lbrace i(I_{11} \xi^2_1+ I_{22}
\xi^2_2 + I_{12} \xi_1\xi_2 + I_{10} \xi_1 + I_{20}\xi_2)\right. \nonumber\\
&+& ( \left.R_{11} \xi^2_1 + R_{22} \xi^2_2 + R_{12} \xi_1\xi_2 +
R_{10} \xi_1 + R_{20} \xi_2)\right\rbrace .
\end{eqnarray}
The expressions of the constants $I_{ij}$ and $R_{ij}$ in terms of
the quantities $ \xi_k, \sigma_k , A_i,B_i $ are given in the
appendix. The preceding formula can be identified with
Eq.(\ref{145}) only if we can make all the $I_{kl}$ vanish as well
as $R_{12}$. The constants $B_1$ and $B_2$ are easily fixed  by
the requirement that
\begin{equation}
\label{148} I_{10} = {B_2\alpha_2+B_1\beta_2\over{\sqrt{\theta}}}
+ {\beta_1\lambda_1+\alpha_1\lambda_2 \over{\theta}} = 0 \quad ,
\end{equation}
and
\begin{equation}
\label{149} I_{20} = {B_2\gamma_2+B_1\delta_2\over{\sqrt{\theta}}}
+ {\delta_1\lambda_1+\gamma_1\lambda_2 \over{\theta}} = 0 \quad .
\end{equation}
In a similar way, the vanishing of $I_{11}$ and $I_{22}$ fix $A_1$
and $A_2$. Simplifying, the remaining coefficients assume the
following expressions
\begin{eqnarray}
\label{150}
I_{12} &=& {a+d\over{2(a+ac+cd)}}{1\over {\theta}} \quad , \\
\label{151} R_{11} &=& {-a-2ac-ac^2+d-c^2d+ad^2+d^3\over
{4(-ac-2ac^2-ac^3+a^2d+a^2cd-c^2d-c^3d+acd^2)}} {1\over {\theta}}  \quad , \\
\label{152} R_{22} &=& {-a^3+2ac+ac^2-a^2d+c^2d\over
{4(-ac-2ac^2-ac^3+a^2d+a^2cd-c^2d-c^3d+acd^2)}} {1\over {\theta}}  \quad , \\
\label{153}
R_{12} &=& {a-d\over{2(-c-c^2+ad)}} {1\over {\theta}} \quad , \\
\label{154}
R_{10} &=& {-(1+c)\lambda_1-d\lambda_2\over {c(1+c)-ad}} {1\over {\theta}} \quad , \\
\label{155} R_{20} &=&
{-(a\lambda_1+c\lambda_2)\over{(-c(1+c)+d)}} {1\over {\theta}}
\quad .
\end{eqnarray}
The remaining term in the imaginary part vanishes only if $ d =
-a$ as is manifest in Eq.(\ref{150}). The $R_{kl}$ coefficients
then simplify further :
\begin{eqnarray}
\label{156} R_{11} &=& {a(1+c)\over{2a(-a^2-c-c^2)}}{1\over
{\theta}}\quad , \quad
R_{12} = {a\over{-a^2-c(1+c)}} {1\over {\theta}} \quad , \quad \\
\label{157} R_{22} &=& {ac\over{2a(-a^2-c-c^2)}} {1\over {\theta}}
\quad  .
\end{eqnarray}
One sees that when $R_{12} = 0, R_{11}$ and $R_{22}$ are only
defined by their limits as $a\rightarrow 0$:
\begin{eqnarray}
\label{158} R_{11} &=& -{1\over {2c}} {1\over {\theta}}\quad ,
\quad R_{22} = -
{1\over {2(1+c)}} {1\over {\theta}} \quad , \\
\label{159} R_{10} &=& {\lambda_1\over {c\theta}}\quad , \quad
R_{20} = {\lambda_2 \over{(1+c)\theta}} \quad .
\end{eqnarray}
In summary, when $d =-a=0$, the choice  of the constants
$A_1,A_2,B_1,B_2$ explained earlier leads to a maximally localized
state which takes the form
\begin{equation}
\label{160} \psi^{ml}_{\lambda_1,\lambda_2} (\xi_1,\xi_2) = N \exp
\left\lbrace {\xi^2_1\over {2c\theta}}
-{\xi^2_2\over{2(1+c)\theta}} - {\lambda_1\xi_1\over{c\theta}} +
{\lambda_2\xi_2\over{(1+c)\theta}}\right\rbrace \quad .
\end{equation}
This state is normalizable only if the  quadratic terms are
negative and this is realized provided that  the constant $c$
assumes values in the interval $ ]-1,0[$. One finally obtains that
the wave function
\begin{equation}
\label{161} \psi^{ml}_{\lambda_1,\lambda_2}(\xi_1,\xi_2) =
 {1\over {\pi \theta}} \frac{1}{\sqrt{-2 c}} \frac{1}{\sqrt{2 (c+1)}} \exp \left\lbrace {1\over
{2c\theta}} (\xi_1-\lambda_1)^2-{1\over {2(1+c)\theta}}
(\xi_2-\lambda_2)^2\right\rbrace
\end{equation}
represents a maximally localized state in the representation of
the non commutative quantum mechanics given by
\begin{eqnarray}
\label{162}
\hat x_1 &=& \xi_1 + i \, \theta \, (1+c) \, \partial_{\xi_2} \quad , \nonumber\\
\hat x_2 &=& \xi_2 + i \, \theta \, c \, \partial_{\xi_1} \quad .
\end{eqnarray}
It is straightforward to verify, by the computation of integrals
implying gaussians multiplied by polynomials that in this state
\begin{eqnarray}
\label{163} \langle x_i \rangle = \lambda_i  \quad   ,  \quad \Delta x_i
= \sqrt{\theta\over 2}  \quad .
\end{eqnarray}
  This ensures that the state fulfills the condition not only in the
  limiting case $ a \rightarrow 0 $, but also in the case $ a = 0 $
  itself. The norm of this state  is perfectly finite
 \begin{equation}
   \langle \psi \vert \psi \rangle = -\frac{1}{c(1+c)} \frac{\pi}{8
   \theta} \quad .
  \end{equation}
However, it goes to infinity as $\theta $
  goes to zero. This agrees with the fact that the square of the Dirac distribution
   is not a mathematically well defined object.
Like in the K.M.M  theory \cite{LABEL9}, one finds that the mean
momentum vanishes in this state
\begin{equation}
\label{164} \langle p_i \rangle = 0 \quad .
\end{equation}
The uncertainty in momentum reads
\begin{equation}
\label{165} \Delta p_i = \left( -{\hbar^2\over{2c\theta}}
\right)^{1\over 2} \quad ,
\end{equation}
 and, along any direction, one has
\begin{equation}
\label{166} \Delta x_i \Delta p_i = {1\over 2}
{\hbar\over{(-c)^{1\over 2}}} \quad .
\end{equation}
We can not  reach the lowest  values allowed by the Heisenberg
uncertainty relations since this corresponds to the value $c=-1$
which blows up  the maximally localized state(see Eq.(\ref{161})).

\par It is quite surprising  that the condition $ a = d = 0 $
   which was needed to recapture some behaviours found using the
   Weyl-Moyal correspondence in the last section is also the one
   leading to gaussians for the maximally localized states. The
   condition $ c = -1/2 $ leads to a  symmetric form of the
   maximally localized states in the variables $ \xi_1, \xi_2 $.
   This strongly suggests a way for the recovering of
 information on position from the Moyal-Weyl wave function.

 \subsection{The quasi-position representation}

\par One can construct a new representation  by
     projecting on the maximally localized states \cite{LABEL9}:
\begin{equation}
   \tilde{\alpha}(\xi_1,\xi_2) = \int d \lambda_1 d\lambda_2 \quad  \psi^{ml}_{\lambda_1, \lambda_2}(\xi_1,\xi_2) \quad
 \alpha(\lambda_1,\lambda_2)
   \quad .
\end{equation}
For simplicity, we restrict ourselves to the case $ c = -1/2 $.
The action of the position operators is given by
\begin{equation}
\label{op} \hat{x}_1 = \lambda_1 + \frac{\theta}{2}(
\partial_{\lambda_1} + i
\partial_{\lambda_2}) \quad , \quad
\hat{x}_2 = \lambda_2 + \frac{\theta}{2}( \partial_{\lambda_2} - i
\partial_{\lambda_1}) \quad ,
\end{equation}
 while the scalar product reads
\begin{eqnarray}
 \langle \tilde{\alpha} \vert \tilde{\beta} \rangle & = & \frac{1}{8 \pi \theta} \int d
 \lambda_1\,
d \lambda_2 \, d \mu_1 \, d \mu_2 \, \tilde{\alpha}^{*}(\lambda_1,
\lambda_2) \, \tilde{\beta}(\mu_1, \mu_2) \nonumber\\
& & \exp{ \left( - \frac{1}{2 \theta}(\lambda_1 - \mu_1)^2 -
\frac{1}{2 \theta}(\lambda_2 - \mu_2)^2 \right) } \quad .
\end{eqnarray}
As $\theta \rightarrow 0 $, the gaussian function tends to the
delta distribution. This enables one to rewrite the preceding
integral as based on two coordinates rather than four, recovering
the well known situation of the position representation. The
operators given in Eq.(\ref{op}) also reduce to the common form in
the same limit.
\par This representation is physically interesting because within it
we know the interpretation of the wave function. Since it is
obtained
   by the projection of maximally localized states, the square of
   the wave function $ \tilde{\phi}(\xi_1,\xi_2)$ gives the
   probability for a particle to be localized in the
   intervals  $ [ \xi_1 -  \frac{1}{2}  \sqrt{\frac{\theta}{2}}, \xi_1 +  \frac{1}{2}  \sqrt{\frac{\theta}{2}}]$
on the first axis and a similar one centered around $ \xi_2$ for the second
   axis.
\par The quasi position representation obtained here is similar to
the one found in the K.M.M theory  in the sense that the scalar
product involves functions defined at different points
\cite{LABEL9}. It differs from it by the fact that the operators
are not given here by an infinite series in the deformation
parameter.

 \subsection{The momentum representation}

  We worked in a representation which reduces to the
 position one in the undeformed limit. It is possible to carry  similar
 calculations in the momentum representation. The parameterization
 \begin{equation}
 \hat{x}_1 = i \partial_{p_1} - \frac{\theta}{2} p_2  \quad ,
 \quad
 \hat{x}_2 = i \partial_{p_2} + \frac{\theta}{2} p_1
 \end{equation}
 satisfies  the commutation relations. One can construct  maximally localized states which tend to $\exp{(i p
 x)}$ which is the Fourier transform of the Dirac delta.

\subsection{Higher dimensions}

\par The construction presented needs some modifications when addressing higher dimensions.
 Let us consider for example a $3+1$  dimensional model whose non vanishing commutation
 relations are
\begin{equation}
\label{167} [\hat x_1, \hat x_2] = [\hat x_2, \hat x_3] = [\hat
x_3, \hat x_1] = i \theta \quad .
\end{equation}
It is realized by the following operators
\begin{eqnarray}
\label{168} \hat x_1 &=& \xi_1+i\theta \, (a_1 \, \partial_{\xi_1}
+ (1+b_1)\, \partial_{\xi_2}
+ a_3 \, \partial_{\xi_3}) \quad , \nonumber\\
\hat x_2 &=& \xi_2+i\theta (b_1 \, \partial_{\xi_1} + b_2 \,
\partial_{\xi_2}
+ (1+c_2) \, \partial_{\xi_3}) \quad , \nonumber\\
\hat x_3 &=& \xi_3+i\theta \, ((1+a_3)\, \partial_{\xi_1} + c_2 \,
\partial_{\xi_2} + c_3 \, \partial_{\xi_3}) \quad .
\end{eqnarray}
The $a_i,b_i,c_i$ are arbitrary but real constants. We want $ x^2
+ y^2 + z^2 $ to be quadratic in the ``momenta" and linear in the
`` angular momenta" like in Eq.(\ref{126}) which is true for all
dimensions. One needs the relations $ a_1 = b_2 = c_3 = 0 $ to
cancel terms of the form $ \xi_k
\partial_{\xi_k} $ and we impose
 $ b_1 = a_3 = c_2 = - \frac{1}{2} $ to ensure the appearance of the ``angular momenta".
We end up with the following expressions for the position operators:
\begin{equation}
\label{op3}
 \hat x_1 = \xi_1+i \frac{\theta}{2} \, \left(  \partial_{\xi_2}
- \partial_{\xi_3} \right) \quad , \quad
 \hat x_2 = \xi_2+i \frac{\theta}{2} \, \left(  \partial_{\xi_3}
- \partial_{\xi_1} \right) \quad , \quad
 \hat x_3 = \xi_3+i \frac{\theta}{2} \, \left(  \partial_{\xi_1}
- \partial_{\xi_2} \right) \quad .
\end{equation}

Inspired by what we have done in the previous subsection, we now look
for the coefficients $\sigma_1$ which allow the function
\begin{eqnarray}
\label{169}
& & \psi^{ml}_{\lambda_1,\lambda_2,\lambda_3}(\xi_1,\xi_2,\xi_3) =
 \left(\frac{1}{2 \pi \theta} \right)^{3/2} \frac{1}{\sigma_1 \sigma_2 \sigma_3}\nonumber\\
& & \exp \left\lbrace -{(\xi_1-\lambda_1)^2\over {\sigma_1^2
\theta}} -{(\xi_2-\lambda_2)^2\over{\sigma_2^2 \theta}}
-{(\xi_3-\lambda_3)^2\over {\sigma_2^2 \theta}}\right\rbrace
\end{eqnarray}
to satisfy the definition of the maximally localized state given
in the fourth section. Using the representation specified in
Eq.(\ref{op3}) and the three dimensional version of the scalar
product given in Eq.(\ref{131}) the conditions on the mean values
of the positions are automatically fulfilled. The ones concerning
the uncertainties lead to the following conditions
satisfied

\begin{equation}
\label{del1}
 \Delta x_1 \Delta x_2 = \frac{\theta}{2} \quad \Longrightarrow  \quad \frac{ \sigma_1^2}{4} +
  \frac{1}{4\sigma_2^2} +
  \frac{1}{4\sigma_3^2} = 0 \quad ,
\end{equation}
\begin{equation}
\label{del2}
  \Delta x_1 \Delta x_3 = \frac{\theta}{2} \quad \Longrightarrow  \quad \frac{1}{4\sigma_1^2} +
  \frac{ \sigma_2^2}{4} +
  \frac{1}{4\sigma_3^2} = 0 \quad ,
\end{equation}
\begin{equation}
\label{del3}
  \Delta x_2 \Delta x_3 = \frac{\theta}{2} \quad \Longrightarrow \quad  \frac{1}{4\sigma_1^2} +
  \frac{1}{4\sigma_2^2} +
  \frac{1 \sigma_3^2}{4} = 0 \quad .
\end{equation}
  We can use  Eqs.(\ref{del1},\ref{del2}) to express $ \sigma_1$ and $
  \sigma_3$ in terms of $ \sigma_2$:
  \begin{equation}
\sigma_1^2 = 2 - \frac{1}{\sigma_2^2} -
     \frac{4 \sigma_2^4}{ 1+ \sigma_2^4 - 2 \sigma_2^2- \sqrt{ 1-4 \sigma_2^2+6 \sigma_2^4 + 4 \sigma_2^6+
      \sigma_2^8}}
       \quad ,
\end{equation}
\begin{equation}
  \sigma_3^2 = \frac{1}{4} + \frac{1}
      {4\sigma_2^4} -
     \frac{1}{2\sigma_2^2} -
     \frac{{\sqrt{1 - 4\sigma_2^2 +
           6\sigma_2^4 +
           4\sigma_2^6 +
           \sigma_2^8}}}{4
        \sigma_2^4} \quad .
\end{equation}
Replacing, in the second part of Eq.(\ref{del3}), the variables $ \sigma_1 $ and $\sigma_3 $ by the expressions obtained in the two preceding formulas, one obtains an equation in the parameter $\sigma_2$.
Unfortunately, this  equation does not admit a real solution, as a
numerical treatment shows. One may think that a different choice
of the parameters $a_i,b_i,c_i$ may solve the problem; this is not
the case.
\par The simplest way to understand this feature is the
following. As shown in
Eq.(\ref{139}), the equality given in Eq.(\ref{8}) for the directions $x_1,x_2$ is satisfied only by
the eigenstates of the operator
\begin{equation}
 \O_{12} = \hat{x_1} + i \hat{x_2} \quad .
\end{equation}
So, a simultaneous solution of Eq.(\ref{8}) for
all couples of directions must be an eigenstate of the operators
$\O_{12},\O_{13}$ and $\O_{23} $. The commutators of these operators are non vanishing:
\begin{equation}
\label{com}
[ \O_{12},\O_{13}] =  2  \theta \quad .
\end{equation}
As we just pointed out, a state which saturates the three bounds is an eigenstate of the three operators; we denote its eigenvalues by $ \lambda_{12},  \lambda_{13}$ and $  \lambda_{23} $. Using this, we infer from Eq.(\ref{com}) the equation
\begin{equation}
[ \O_{12},\O_{13}] \vert \psi \rangle = 2  \theta \vert \psi
\rangle = ( \lambda_{12} \lambda_{13} - \lambda_{13} \lambda_{12})  \vert \psi
\rangle = 0 \quad .
\end{equation}
so  that the only
state which saturates all the three bounds is the null vector of
the Hilbert space {\sl composed of the appropriate functions}.
This means that we can not saturate the three bounds by states
{\sl which are functions} for the geometry displayed in
Eq.(\ref{167}). Solving  Eq.(\ref{del1}), one
saturates the bound $ \Delta x_1 \Delta x_2$; the non commutation
of the operators forbids one to simultaneously satisfy the same
relation for $ \Delta x_2 \Delta x_3 $. This is the explanation of
the failure to implement simultaneously the set of equations displayed in
Eq.({\ref{del1},{\ref{del2},{\ref{del3}).

\par The construction of the preceding subsections can however
prove useful. Let us consider the situation in which non locality
is confined to the plane $x_1,x_2$:
\begin{equation}
[ \hat{x}_1, \hat{x}_2 ] = i \theta, [ \hat{x}_1, \hat{x}_3] = [ \hat{x}_2, \hat{x}_3 ] = 0 \quad .
\end{equation}
As the third direction does commute with the others, there is no
Heisenberg uncertainty relation which prevents us from taking $
\Delta x_3 = 0$. The situation in the non commuting plane is
exactly the one studied in the previous section. Taking a
representation in which the supplementary operators are given by $
\hat{x}_3 = \xi_3 , \hat{p}_3 = - i \hbar
\partial_{\xi_3}$, a maximally localized state can then be obtained as the
product {\sl of a function and a delta distribution}:
\begin{equation}
\psi^{ml}_{\lambda_1,\lambda_2}(\xi_1,\xi_2) =
 {1\over {\pi \theta}} \frac{1}{\sqrt{-2 c}} \frac{1}{\sqrt{2 (c+1)}} \exp \left\lbrace {1\over
{2c\theta}} (\xi_1-\lambda_1)^2-{1\over {2(1+c)\theta}}
(\xi_2-\lambda_2)^2\right\rbrace  \delta{(\xi_3 - \lambda_3)} \quad
.
\end{equation}

\section{ Causality of a Q.F.T with a non commuting time}\label{sec_causa}

   The analysis of \cite{LABEL6}  which led to the
   conclusion that quantum field theories with a non commuting
   time were acausal relies on the interpretation of the wave
   function   as giving the probability amplitude. This is valid in the
   ordinary
    ( $ \theta = 0 $  ) theory, but when non commutativity sets in, one can not simultaneously
    diagonalize  the  coordinates. The Heisenberg-like uncertainty
      forbids one
     to speak of an event happening at a time and a place known
     with infinite precision.
        What can we do to gain information on time and position in
        this context? The useful procedure was developed by
        K.M.M \cite{LABEL9} in a different model : it is the projection on maximally
        localized states.

\par To give an idea of how our analysis may alter the causality
issue, let us summarize  the analysis of \cite{LABEL6}.
   The theory under study is  two dimensional and invariant under the Lorentz group:
\begin{equation}
\label{178}
 [\hat x_{\mu}, \hat x_{\nu}] = i \theta
\epsilon_{\mu\nu} .
\end{equation}
In  \cite{LABEL6}, the ``time" coordinate is written $t$ and the
``space" coordinate $x$. One Considers an incoming state of
correlated pairs of
     particles with opposite momenta
\begin{equation}
\label{182}
  \mid \phi \rangle_{in} = \int \frac{dk}{( 2 \pi) 2 E_k } \phi_{in}(k)
  \mid k ,-k \rangle  \quad ,
  \end{equation}
  centered at two momenta $ po$ and $ - po$:
 \begin{equation}
 \label{184}
 \phi_{in}(p) = E_{p} \left( \exp{ \left( - \frac{(p - po)^2}{\lambda} \right)}
+ \exp{\left( - \frac{(p + po)^2}{\lambda} \right)} \right) \quad
,
\end{equation}
with $ E_{p} = \sqrt{ p^2 + m^2 }$. One has that at  ``times" $ t
< 0$, the two packets are well separated. At $ t = 0 $, the wave
function is  concentrated at the ``position" $x = x_1 - x_2 = 0 $(
$ x_i$ is the mean ``position" of the i th wave packet). Then  a
collision takes place, due to the interaction. Considering a final
state of the form
\begin{equation}
\label{1840}
  \mid \phi \rangle_{out} = \int \frac{dp}{( 2 \pi) 2 E_p } \phi_{out}(p)
  \mid p ,-p \rangle \quad ,
  \end{equation}
( $ \phi_{out}(p)$ is related to $ \phi_{in}(p)$ by the $S$
matrix) it is found that for the usual $ \phi^4 $ theory, the
outgoing wave function simply displays a
    small time delay.
The corresponding non commutative theory(i.e. with the
interaction term $ g \, \phi * \phi * \phi
* \phi $ using the Moyal product)behaves very differently. The wave packet, written in the ``position" space,
displays three peaks.  The
   last peak leaves the collision point $ x = 0 $  before the incoming wave packets
   given in Eq.(\ref{184})arrive there and this is interpreted  as a violation
   of causality. The interested reader will find all the details in \cite{LABEL6}

 \par As we emphasized at the end of the second section and in the third one, the
 $x,t$ variables
 appearing in the Moyal product are mere notations; they coincide
 with the physical position and time only when $ \theta = 0 $. To obtain viable informations on
  positions, one has to project on
     maximally localized states.
{\sl In these theories, a
 phenomenon
  can not occur at a perfectly known time and a perfectly known
  position}. Any instant, any position is surrounded by a zone of fuzziness.
 Then, a more careful formulation should lead to a situation in which
  the collision between the ingoing packets is described as taking place during the time
interval $ [ \lambda_{i0} - \frac{1}{2} \sqrt{\frac{\theta}{2}},
\lambda_{i0} + \frac{1}{2} \sqrt{\frac{\theta}{2}} ]$ in the
position interval $[ \lambda_{i1} - \frac{1}{2}
\sqrt{\frac{\theta}{2}}, \lambda_{i1} + \frac{1}{2}
\sqrt{\frac{\theta}{2}}]$. The third outgoing packet will leave
the region of collision at a time lying in an interval
$[\lambda_{f0} - \frac{1}{2} \sqrt{\frac{\theta}{2}},\lambda_{f0}
+ \frac{1}{2} \sqrt{\frac{\theta}{2}}]$ . If these two time
intervals are not disjoints, one can not speak of an acausal
process because of the fuzziness concerning time. The critical
point concerns the calculation of the instants $ \lambda_{i0}$ and
$ \lambda_{f0}$ which we do not have for the time being. One has
to be especially careful since the status of the position
operators in Q.F.T  is not exactly the one present in quantum
mechanics. In the commutative case, this is embodied in the fact
that the appropriate Newton-Wigner position operators are not
simply the derivatives of the fields in the momentum
representation \cite{LABELsil}. So, one needs more to draw a
definite conclusion .

\par  The promising point in this
picture is that the motion of the two incoming ``particles" is not
symbolized by two lines in the  time-position  plane but by two
ribbons.
\par  Nevertheless, some
technical and conceptual problems must be addressed before a more
elaborate treatment. One has  to understand  the hamiltonian
structure of the theories with non commuting time better: the conjugate
of the field is an infinite series, the energy momentum tensor and
the current are not conserved, etc.

\section{conclusion}\label{sec_conc}

\par In this work, we have shown how some results obtained using
the Moyal product can be recovered by the choice of a particular
representation of the position and momentum operators. We have
shown that the maximally localized states associated to these
representations can be chosen to be  gaussian functions which tend to the delta
distribution as the parameter of non commutativity is sent to
zero. We have also suggested how these states may alter  the
analysis of the causality issue of a  theory in which space and
time do not commute.

\par One may ask if the representation we choosed reproduces the
results which can be obtained using the Moyal product for any
physical system. We do not know the answer yet.
\par The method we used here to study non commutative quantum
    mechanics is closer to \cite{LABEL9} than to \cite{LABEL14} in
    the sense that we did not introduce a differential
    calculus compatible with the commutation relations between the
    coordinates. This structure is usually used to construct an invariant action
    which leads to the field equations. Our procedure may  not be applicable to
      curved spaces, contrary to the method used in
      \cite{LABEL14}.
\par It should be noted that the non commutation of the positions
raises  a supplementary ordering problem. We did not face it because we studied
central potentials only.

\bigskip
\underline{Acknowledgement} I  thank J.Madore for encouraging
conversations at \mbox{Orsay} during my stay at I.A.P. in Paris
and I thank the F.N.R.S for its financial support . I also thank
Ph.Spindel for his advices.

\appendix
\section{Coefficients}\label{AppA}

  We give the explicit form of the coefficients appearing in Eq.(\ref{147}).
  \begin{eqnarray}
  \label{A.1}
  I_{11} &=&  {1\over {\theta}}
\left(-{a \alpha_1^2\over 2} + {d\alpha_1^2\over 2} -\alpha_1
\alpha_2 + A_2\alpha_2^2 + \alpha_1\beta_1 + 2 c\alpha_1\beta_1
- a \alpha_2 \beta_1- d\alpha_2 \beta_1 \right.\nonumber\\
&+& \left.{a\beta_1^2\over 2} - {d\beta_1^2 \over 2} - a \alpha_1
\beta_2 - d \alpha_1 \beta_2 + 2 A_1 \alpha_2 \beta_2 + \beta
1\beta_2 - A2\beta_2^2\right) \quad .
\end{eqnarray}
\begin{eqnarray}
\label{A.2} I_{22} &=& {1\over {\theta}} \left(-{a\gamma_1^2\over
2} + {d\gamma_1^2\over 2} - \gamma_1 \gamma_2+ A2 \gamma_2^2
+ \gamma_1\delta_1+ 2 c \gamma_1 \delta_1 - a\gamma_2\delta_1\right.\nonumber\\
&-& \left.d\gamma_2\delta_1+{a\delta_1^2\over 2} - {d\delta_1^2
\over 2} - a \gamma_1 \delta_1 \right.\nonumber\\
&-& \left. d\gamma_1\delta_2+ 2 A_1\gamma_2 \delta_2 + \delta_1
\delta_2-A_2\delta_2^2\right) \quad .
\end{eqnarray}
\begin{eqnarray}
\label{A.3}  I_{12} &=& {1\over {\theta}}  \left(-a\alpha_1
\gamma_1 + d \alpha_1 \gamma_1- \alpha_2 \gamma_1 + \beta_1
\gamma_1 + 2c \beta_1
\gamma_1-a \beta_2\gamma_1- d \beta_2 \gamma_1 - a_1\gamma_2 \right.\nonumber\\
 &+& \left. 2 A_2 \alpha_2 \gamma_2 - a \beta_1 \gamma_2- d \beta_1
\gamma_2 + 2 A_1 \beta_2 \gamma_2 + \alpha_1 \delta_1 + 2 c
\alpha_1 \delta_1 - a \alpha_2 \delta_1 - d \alpha_2
\delta_1 \right.\nonumber\\
&+& \left. a \beta_1 \delta_1 - d\beta_1 \delta_1 + \beta_2
\delta_1 - a \alpha_1 \delta_2- d \alpha_1 \delta_2 + 2 A_1
\alpha_2 \delta_2 + \right.\nonumber\\
& & \left. \beta_1 \delta_2-2 A_2 \beta_2 \delta_2\right) \quad .
\end{eqnarray}

\begin{eqnarray}
\label{A.6} R_{11} &=& {1\over {\theta}}  \left({\alpha_1^2\over
2} + c \alpha_1^2 - a \alpha_1 \alpha_2 - d\alpha_1 \alpha_2
+ A_1 \alpha_2^2 + a \alpha_1 \beta_1 -d \alpha_1 \beta_1 + \alpha_2\beta_2\right.\nonumber\\
&-& \left. -{\beta_1^2\over 2} - c\beta_1^2 + \alpha_1\beta_2 -
2A2 \alpha_2 \beta_2 + a \beta_1\beta_2 + d\beta_1\beta_2-
A_1\beta_2^2\right) \quad .
\end{eqnarray}

\begin{eqnarray}
\label{A.7} R_{22} & = & {1\over {\theta}}  \left({\gamma_1^2\over
2} + c \gamma_1^2 - a \gamma_1 \gamma_2 - d\gamma_1 \gamma_2
+ A_1\gamma_2^2 + a \gamma_1\delta_1-d\gamma_1\delta_1 + \gamma_2\delta_1\right.\nonumber\\
&-& \left. -{\delta_1^2\over 2} - c\delta_1^2 + \gamma_1\delta 2 -
2A_2 \gamma_2 \delta_2 + a \delta_1\delta_2 + d\delta_1\delta_2-
A_1\delta_2^2\right) \quad .
\end{eqnarray}
\begin{eqnarray}
\label{A.8} R_{12} & = & {1\over {\theta}}  \left( -a\alpha_1
\gamma_1 + 2c \alpha_1 \gamma_1- a\alpha_2 \gamma_1 + a\beta_1
\gamma_1 -d \beta_1
\gamma_1+ \beta_2\gamma_1- a \alpha_1 \gamma_2-d\alpha_1\gamma_2 \right.\nonumber\\
&+& \left. + 2A_1 \alpha_2\gamma_2 + \beta_1 \gamma_2 - 2A_2
\beta_2 \gamma_2 + a \alpha_1 \delta_1 - d \alpha_1 \gamma_1 +
\alpha_2 \delta_1-\beta_1\delta_1 - 2c\beta_1\delta_1
\right.\nonumber\\
&+& \left. a \beta_2 \delta_1 + d \beta_2 \delta_1 + \alpha_1
\delta_2 - 2A_2 \alpha_2 \delta_2 + a \beta_1 \delta_2 + d \beta_1
\delta_2 - 2 A_1 \beta_2 \delta_2 \right) \quad .
\end{eqnarray}
\begin{equation}
\label{A.9} R_{10}  = {B_1 \alpha_2 - B_2 \beta_2\over
{\sqrt{\theta}}} + {\alpha_1 \lambda_1 - \beta_1\lambda_2\over
{\theta}} \quad .
\end{equation}
\begin{equation}
\label{A.10} R_{20} ={B_1\gamma_2 - B_2 \delta_2\over
{\sqrt{\theta}}} + {\gamma_1 \lambda_1 - \delta_1\lambda_2\over
{\theta}} \quad .
\end{equation}


\begin{references}
\bibitem{LABEL1}
N.Seiberg, E.Witten JHEP 9909(1999)032.
\bibitem{LABEL2}
D.Bigatti, L.Susskind Phys.Rev.D.{\bf{62}},(2000)066004.
\bibitem{LABEL3}
S.Coleman, S.L.Glashow Phys.Rev.{\bf{D.59}}(1999)116008.
\bibitem{LABEL4}
F.W.Stecker, S.L.Glashow, astro-ph/0102226.
\bibitem{LABEL6}
N.Seiberg, L.Susskind,N.Toumbas {\bf JHEP} 0006(2000)044.
\bibitem{LABEL7}
M.Chaichian, A.Demichev, P.Presnajder,M.M.S.Jabbari, A.Tureanu,
hep-th/0101209.
\bibitem{LABEL8}
M.Chaichian, A.Demichev, P.Presnajder, Nucl.phys.B567(2000)360,
J.Math.Phys.{\bf 41}(2000)185.
\bibitem{LABEL9}
\noindent A. Kempf, G. Mangano, R.B. Mann,
Phys.Rev.D{\bf{52}},(1995)1108. \newline H.Hinrichsen, A.Kempf,
J.Math.Phys.{\bf 37}(1996)2121. \newline S.L.Adler , A.Kempf ,
J.Math.Phys.{\bf 39}(1998)5083.
 \newline A.Kempf, S.Majid, J.Math.Phys.{\bf 35}(1994)6802.
 \newline A.Kempf, J.Math.Phys.{\bf 35}(1994)4483.
\bibitem{LABELD} S.Doplicher, Comm.Math.Phys(1995)187.
\bibitem{LABEL10}
 J.Gamboa, M.Loewe, J.C.Rojas, hep-th/0010220.
\bibitem{LABEL11}
L.Mezincescu, hep-th/0007046.
\bibitem{LABEL12}
R.Brout, Cl.Cabriel,M.Lubo, Ph.Spindel, Phys.Rev. {\bf D59} (1999)
044005.
\bibitem{LABEL13}
M.Lubo,  Phys.Rev.{\bf D61}(2000)124009.
\bibitem{LABEL14}
S.Chow, R.Hinterding,J.Madore,H.Steinacker, LMU-TPW
99-06,MPI-PhT/99-12.
\bibitem{LABELsil}
Silvan S.Schweber,{\sl An Introduction To Relativistic Quantum
Field Theory}(Harper International edition, New York,U.S.A
$1966$).
\end{references}
\end{document}